\documentstyle[aps,preprint,prb,eqsecnum]{revtex}

\draft

\tighten

\begin{document}

\title{ Coulomb effects on the quantum transport of a two-dimensional 
electron system in periodic electric and magnetic fields.} 
\author {Andrei Manolescu}
\address{
Institutul de Fizica \c{s}i Tehnologia Materialelor, C.P. MG-7
Bucure\c{s}ti-M\u{a}gurele, Rom\^ania}
\author{Rolf R. Gerhardts}
\address{
Max-Planck-Institut f\"ur Festk\"orperforschung, Heisenbergstrasse 1,
D-70569 Stuttgart, Federal Republic of Germany}

\maketitle

\begin{abstract}

The magnetoresistivity tensor of an interacting two-dimensional electron 
system with a lateral and unidirectional electric or magnetic modulation, 
in a perpendicular quantizing magnetic field, is calculated within the 
Kubo formalism.  The influence of the spin 
splitting of the Landau bands and of the density of states (DOS) on the 
internal structure of the Shubnikov-de Haas oscillations is analyzed.  The 
Coulomb electron-electron interaction is responsible for strong screening 
and exchange effects and is taken into account in a screened Hartree-Fock 
approximation, in which the exchange contribution is calculated 
self-consistently with the DOS at the Fermi level.  This approximation 
describes both the exchange enhancement of the spin splitting and 
the formation of
compressible edge strips, unlike the simpler Hartree and Hartree-Fock 
approximations, which yield either the one or the other.

\end{abstract}



\section{Introduction}

Modern techniques allow the fabrication of semiconductor
heterostructures incorporating a two-dimensional electron gas (2DEG) and a 
lateral periodic electrostatic potential (electric modulation) and/or a 
periodic magnetic field (magnetic modulation). In the presence of an external, 
constant, and perpendicular magnetic field, the modulation lifts the 
degeneracy of the Landau levels.  The resulting Landau-band structure 
determines various oscillations of the magnetoresistivities, which usually
provide with the only accessible information on the modulation strength. 

There are at least three types of modulation effects on the 
magnetoresistivities.  First, the Weiss commensurability oscillations in 
the quasi-classical regime of low magnetic fields, both for the electric,
\cite{WKP1} and, more recently for the magnetic modulations,\cite{CGN,YWG} 
have attracted most of the experimental and theoretical work.\cite{PV,G1}
Second, the peculiar subband structure generated by a two-dimensional
superlattice, known as the Hofstadter butterfly, leads to another type of 
commensurability oscillations, inside the Shubnikov-de Haas (SdH) peaks,
their observation being currently the aim of important efforts.\cite{SEKH}  
And third, at high magnetic fields, the profile of the density of states 
(DOS) associated to the energy dispersion of the Landau bands, together with 
the exchange-enhanced spin splitting, may also determine an internal 
structure of the SdH maxima. In the present paper we will discuss only this 
last type of effects.  

The experimental results have been obtained for the resistivity $\rho_{xx}$ 
of GaAs-AlGaAs interfaces with an electric modulation in the $x$ direction, 
i.e. of a one-dimensional character.   The first investigation was performed 
on a modulation created by holographic illumination.\cite{WKP2}
In the absence of the modulation the spin splitting of the second Landau 
level, with $n=1$, could not be resolved in the magnetoresistivity, but with
increasing the modulation amplitude, the evolution of a double-peak 
structure was observed, and it was attributed to the van Hove 
singularities (vHS) of the DOS.

Further experiments, on higher mobility samples modulated by etching 
techniques, have clearly shown the spin splitting. For a weak modulation 
again a double-peak, but also a more complicated triple-peak structure have 
been detected in the spin-polarized SdH maxima of $\rho_{xx}$ corresponding 
to the band $n=1$.\cite{MGT}  For a stronger modulation the spin splitting
vanishes, but the magnetoresistivity {\em minima} at low even filling factors 
change into {\em maxima} at higher even filling factors, simultaneously with 
the  shift of the minima towards the odd filling factors.\cite{TWM}
This behavior has convincingly been explained by the cumulated effects of 
the overlapping vHS from adjacent Landau bands.  In the experiments 
mentioned so far the modulation period has been in the range of 300-500 nm, 
and much larger than the magnetic length.

Another series of recent measurements has been performed on modulated systems 
produced by growing on vicinal surfaces.\cite{SPL}  This technique generates 
an electric modulation of a much shorter period, about 30 nm. The resistivity 
$\rho_{xx}$ also displays a multi-peak structure which may be related to the 
vHS.  The anisotropy of the resistivity, as well as an abrupt onset of the 
spin splitting for a high magnetic field, have been clearly 
evidenced.\cite{HPE}  

In all these experiments a detailed interpretation of the results is 
difficult, and still insufficiently clear. Beyond the technological 
constrictions, the difficulties arise from the complicated relationship 
between the DOS and the magnetoresistivities, and from the non-negligible 
electron-electron interaction effects.  Therefore, a transport calculation
combining on the same footing the electron-modulation, electron-electron 
and the electron-impurity interactions, and also thermal effects, is needed.  

The anisotropy of the modulation may result in a high anisotropy of the 
conductivity tensor.  The conductivity $\sigma_{xx}$ has a scattering
component (inter Landau level), which depends quadratically on the DOS
and {\em vanishes} in the absence of the impurities. The conductivity 
$\sigma_{yy}$ has an additional band component (intra Landau level), 
determined by the dispersion of the one-particle energies, which 
{\em diverges} for weak 
electron-impurity collision broadening.\cite{AV}   Therefore, in a high 
mobility system, the band conductivity may cover the DOS effects in 
$\sigma_{yy}$ and in the related resistivity $\rho_{xx}$.\cite{ZG}

The Coulomb interaction also yields opposite effects. The tendency of the
electrostatic screening is to reduce the energy dispersion imposed by the 
modulation on the effective single-particle states, and hence to {\it 
increase} the DOS.  At the same time, the exchange interaction lowers the 
energy of the occupied states, enhances the energy gaps, but also broadens 
the Landau bands, {\it decreasing} the DOS.  We have recently calculated the 
energy spectra\cite{MG} and the resistivity tensor\cite{MGT} for a
modulated system, in the Hartree-Fock approximation (HFA), and we could 
explain the spin splitting observed in the magnetotransport experiments.  
However, other exchange effects have been overestimated in the standard HFA.  

The main artifact in the HFA results has been the appearance of strong 
short-range charge-density oscillations in the presence of a weak external 
modulation of period much 
longer than the magnetic length, for any filling factor. The reason is the 
competition of the Hartree interaction, of a repulsive character, with the 
Fock interaction, of an attractive character, which in the presence of the 
modulation may excite high charge-density harmonics.\cite{AM3} 
This is reminiscent of the fact that the HFA predicts an instability of the 
homogeneous 2DEG against the formation of a charge density wave for any 
filling factor,\cite{FPA} whereas experiments indicate an inhomogeneous 
ground state (Wigner crystal) only for very low filling factors.
 Another 
consequence of the strong exchange energy is a strong exchange broadening of 
the Landau levels.  A related implication is a substantial, qualitative 
contradiction between the HFA and the results of the Hartree\cite{CSG,BPT} 
or Thomas-Fermi calculations\cite{LG} of edge states.  While the latter 
predict compressible edge strips much wider than the magnetic length, but only 
the bare spin splitting (which is negligible for GaAs), the HFA gives 
considerably narrower compressible 
edge channels, but a strong splitting.\cite{DGH,RB,SB,MG} 
However, the experimental confirmation of wide edge channels\cite{ZHKE}  
suggests the domination of the electrostatic effects.

In order to avoid, or at least to minimize, these artificial features of 
the HFA, our previous attempt to include the Coulomb interaction in a 
magnetotransport calculation for a modulated systems has been limited to 
short modulation periods.\cite{MGT}  The steepness of the energy dispersion, 
on the magnetic length scale, can reduce the relative importance of the 
exchange
interaction, such that the HFA may become qualitatively reasonable. 

In the present paper we want to extend our calculations to the situation
when the modulation period is much longer than the magnetic length.
Therefore our efforts will be mainly focussed on the electron-electron
interaction. Some preliminary results have already been reported.\cite{MGW}  
Our approach is based on a screened HFA (SHFA), in which we include the 
influence of screening on the exchange interaction.  Although we consider
only static screening, this leads already to the desired reduction of
the exchange effects
and avoids the artifacts of the bare HFA.
  The screening is mainly determined by the DOS at the 
Fermi level.\cite{WGG}  Therefore, when the latter will be in 
an energy gap, the screening will be weak, such that the exchange enhancement 
of the Zeeman splitting will remain essentially like in the HFA.   
However, when the Fermi level will intersect a Landau band, the 
exchange effects will substantially diminish.  

The Coulomb interaction is included in a transport calculation, within
the standard Kubo formalism.  For this purpose we need to consider an
electron-impurity scattering mechanism.  We will describe it within
a self-consistent Born approximation (SCBA).

The realization of a magnetic modulation with a period of a few hundred 
nanometers, and a large amplitude, of the order of 1 T is technically 
feasible.\cite{YW}  In the presence of a constant external magnetic field
such a modulation will produce Landau bands and a charge-density response.
Hence, screening and exchange effects will occur, as for an electric 
modulation.  Even in the absence of relevant magetotransport measurements, 
we will include in our calculation such a periodic magnetic field.
For simplicity we will discuss only the case when the modulation, electric 
or magnetic, is unidirectional and sinusoidal. 

The paper is organized as follows.  In Sec.II we derive the 
self-consistent equations of the SHFA which  give us the ground state 
of the system.  Then, in Sec.III, we  discuss the impurity scattering
and the conductivity tensor.  The numerical results of the transport 
calculation are presented in Sec.IV, and the conclusions are 
collected in Sec.V.  Some technical details are given in two
Appendices.

\section{Screened Hartree-Fock Approximation}

We combine the influence of the electron-electron and of the 
electron-impurity interactions on the single-particle states of the 
modulated 2DEG with the help of the average Green function, having the 
operatorial definition
\begin{equation}
G(E)\equiv{\langle{\widehat G}^{-}(E)\rangle}_{imp}=\frac{1}
{E-(H^0+\Sigma^{ee}+\Sigma^{ei}(E))}\,,
\label{gop}
\end{equation}
with the following notations: ${\langle ... \rangle}_{imp}$ stands for the 
average over all the impurity configurations;
$\widehat G^{\pm}(E)=(E-H\pm i0^+)^{-1}$,
with $H$ a generic one-body Hamiltonian of the interacting 2DEG with
impurities; $H^0$ is the Hamiltonian of the noninteracting
2DEG without impurities; $\Sigma^{ee}$ and $\Sigma^{ei}$ are the 
self-energy operators determined by the electron-electron and 
electron-impurity interactions respectively.

In our case the noninteracting Hamiltonian has the form
\begin{equation}
H^0=\frac{1}{2m^*}({\bf p}+e{\bf A(r)})^2+V\cos Kx
-\frac{\sigma}{2}g\,\mu_BB(x).\,
\label{ham0}
\end{equation}
The electrons are located in the plane $\{{\bf r}=(x,y)\}$.  
$B(x)$ is the projection of the magnetic 
field along the $z$ axis and it may have a periodic component, 
$B(x)=B_0+B_{1}\cos Kx$, similar to the periodic electrostatic potential, 
$V(x)=V\cos Kx$.  We choose the vector potential in the Landau gauge, as 
imposed by the symmetry of our system,
\begin{equation}
{\bf A(r)}=(0,B_0x+\frac{B_{1}}{K}\sin Kx). 
\label{vecpo}
\end{equation}
We have also included in $H^0$ the Zeeman term, where $\sigma=+$ for 
spin-up and $\sigma=-$ for spin-down states, $g$ is the bare,
band structure g-factor, and $\mu_B$ the Bohr magneton.

We use the eigenfunctions of $H^0$ corresponding to the unmodulated
system, i.e. for $B_{1}=0$ and $V=0$, as the basis for the one-particle 
Hilbert space.  These functions are the well-known Landau wave functions,
$f_{nX_0}(x,y)=L_y^{-1/2}exp(-iX_0y/l^2)f_{nX_0}(x)$, where $f_{nX_0}(x)$ 
are the one-dimensional harmonic-oscillator wave functions, with
$n=0,1,...$, centered at the position $X_0$, called the center coordinate. 
Here $l$ is the magnetic length, and we will denote by $\omega_c$ the 
cyclotron frequency, both corresponding to the {\it uniform} component of 
the magnetic field $B_0$, $l=(\hbar/eB_0)^{1/2}$ and $\omega_c=eB_0/m^*$.  
In order to simplify the notations we keep the same symbol for the 
two-variable wave function which depends on both spatial coordinates $x$ 
and $y$, e. g. $f_{nX_0}(x,y)$, and for the reduced wave function depending 
only on $x$, $f_{nX_0}(x)$.  The distinction will be made by the number of 
variables specified inside the brackets. The plane-wave factor has been 
normalized to the macroscopic length $L_y$.  The matrix elements of $H^0$ 
are given in the Appendix A.

We will assume randomly distributed impurities, such that the modulated
system is invariant to translations along the $y$ axis.  Consequently,
the dependence on $y$ of the interacting, effective, one-particle wave 
functions, also factorizes in a simple plane wave.  For these wave 
functions we use the notation $\psi_{n\sigma X_0}(x,y)=
L_y^{-1/2}exp(-iX_0y/l^2)\psi_{n\sigma X_0}(x)$, and we expand them 
in the Landau basis,
\begin{equation}
\psi_{n\sigma X_0}(x)=\sum_{n'} c_{nn'}(\sigma,X_0)
f_{n'X_0}(x)\,.
\label{effwf}
\end{equation}
For the interacting, unmodulated system, we have 
$c_{nn'}(\sigma,X_0)= \delta_{nn'}$, and the dependence on the spin label
arises as long as the exchange interaction and an external modulation are
simultaneously present.  The periodic fields broaden the degenerate Landau 
levels $E_{n\sigma}$ into energy bands, $E_{n\sigma X_0}$, which we find 
by solving the eigenvalue problem
\begin{equation}
\left(H^0+\Sigma^{ee}\right) \psi_{n\sigma X_0}(x,y)=
E_{n\sigma X_0} \psi_{n\sigma X_0}(x,y)\,.
\label{scheq}
\end{equation}

The averaged effect of the impurities consists in the spreading 
of the effective single-particle 
energies around the energy spectrum given by Eq.(\ref{scheq}), and thus in 
an additional energy broadening.  The statistical weight of an arbitrary 
energy $E$ is given by the spectral function,
\begin{equation}
\rho_{n\sigma X_0}(E)=\frac{1}{\pi}\,Im\,\langle\psi_{n\sigma X_0}\mid G(E)
\mid\psi_{n\sigma X_0}\rangle \,,
\label{spefu}
\end{equation}
and the contribution of the effective state $(n\sigma X_0)$
to the filling factor $\nu$ can be defined as
\begin{equation}
\nu_{n\sigma X_0}=\int\,dE\,\rho_{n\sigma X_0}(E){\cal F}(E)\,,
\label{ocnum}
\end{equation}
${\cal F}(E)=[exp(E-\mu)/T+1]^{-1}$ being the Fermi function, with $\mu$ 
the chemical potential and $T$ the temperature, such that
\begin{equation}
\nu=\sum_{n\sigma}\int_0^a \frac{dX_0}{a}\,\nu_{n\sigma X_0} \,.
\end{equation}
We will discuss the electron-impurity interaction model in the next section.
Clearly, in the absence of the impurities
$\rho_{n\sigma X_0}(E)\equiv\delta(E-E_{n\sigma X_0})$.

For calculating the electron-electron self-energy we start with the form 
given by the HFA.\cite{FW,DOL}  In the Landau basis the matrix elements  
$\Sigma^{ee}_{n\sigma X_0,n'\sigma 'X'_0}$ do not mix the spin and the
center-coordinate quantum numbers, and can be written as
\begin{eqnarray}
&&\Sigma^{ee}_{nn'}(\sigma,X_0)=\sum_{m\tau Y_0}\nu_{m\tau Y_0}
\int\,d{\bf r}\,d{\bf r'}\,f^*_{nX_0}(x,y)\psi^*_{m\tau Y_0}(x',y')
u({\bf r}-{\bf r'})\nonumber\\
&&\times\left[f_{n'X_0}(x,y)\psi_{m\tau Y_0}(x',y')-\delta_{\sigma\tau}
f_{n'X_0}(x',y')\psi_{m\tau Y_0}(x,y)\right]\nonumber\\
&&\equiv\Sigma^{ee,H}_{nn'}(X_0)+\Sigma^{ee,F}_{nn'}(\sigma,X_0)\,,
\label{hfa}
\end{eqnarray}
where $u({\bf r})=e^2/(\kappa\!\mid{\bf r}\mid)$ is the Coulomb potential 
with $\kappa$ the dielectric constant of the semiconductor background.
The two terms in the square 
brackets of Eq.(\ref{hfa}) define the Hartree (direct) and the Fock 
(exchange) contributions, indicated by the superscripts $H$ and $F$.
The Hartree term can be rewritten in terms of the particle density,
\begin{equation}
n(x)=\sum_{m\tau Y_0}\nu_{m\tau Y_0}\mid\psi_{m\tau Y_0}(x)\mid^2
=\sum_{p\geq 0} n_p \cos\,pKx\,,
\label{parden}
\end{equation}
the last form being the Fourier expansion appropriate to our external 
fields.  The Fourier coefficients of the Hartree self-energy are given in 
Appendix A.  

Using Eq.(\ref{effwf}) and the Fourier transform of the Coulomb potential,
$\tilde u({\bf q})=2\pi e^2/(\kappa\!\mid{\bf q}\mid)$, the Fock self-energy 
becomes
\begin{eqnarray}
&&\Sigma^{ee,F}_{nn'}(\sigma,X_0)=-\sum_{mY_0}\nu_{m\sigma Y_0}
\sum_{m_1m_2}c_{mm_1}(\sigma,Y_0)c_{mm_2}(\sigma,Y_0)\nonumber\\
&&\times\int\,\frac{d{\bf q}}{(2\pi)^2}\tilde u({\bf q})
\langle f_{nX_0}\mid e^{i{\bf qr}}\mid f_{m_1Y_0}\rangle
\langle f_{m_2Y_0}\mid e^{-i{\bf qr}}\mid f_{n'X_0}\rangle\,.
\label{fter}
\end{eqnarray}
The Fourier expansion of $\Sigma^{ee,F}_{nn'}(\sigma,X_0)$ is also given 
in Appendix A.  

In order to overcome the artifacts of the HFA mentioned in the Introduction,
we screen the Coulomb potential in the exchange self-energy, substituting 
$\tilde u({\bf q})$ in Eq.(\ref{fter}) by
\begin{equation}
\tilde u({\bf q})=\frac{e^2}{\kappa}\,\frac{2\pi}{q\,\epsilon(q)} \,,
\label{scrint}
\end{equation}
where $\epsilon(q)$ is the {\it static} dielectric function.
We thus neglect the dynamic screening effects.  In the spirit of the
random-phase approximation, $\epsilon(q)=1-\frac{2\pi e^2}{\kappa q}\chi(q)$, 
$\chi(q)$ being the dielectric susceptibility given by the well-known
Lindhard formula,\cite{FW} 
to be evaluated
self-consistently with the effective states
resulting from Eq.(\ref{scheq}).  In the diagrammatic picture, Fig.1,
that means the series of the polarization loops --- previously contained
only in the Hartree part of $\Sigma^{ee}$ --- is now included, 
via Eq.(\ref{scrint}), also in the Fock part.

For a two-dimensional system in a perpendicular magnetic field one can 
identify two  components of the static dielectric susceptibility,
\begin{equation}
\chi(q)=\chi_1(q)+\chi_2(q) \,,
\label{chi}
\end{equation}
corresponding to the {\it intra-} and to the {\it inter-} Landau level
transitions respectively \cite{GG}.  In other words, $\chi_1(q)$ describes 
the electrostatic response due to the electron redistribution around 
the Fermi level, under the action of an electric field of an arbitrary wave 
vector ${\bf q}$, while $\chi_2(q)$ gives the response due to the distortion 
of the effective wave functions.  

For $\chi_1$ we use the expression derived for {\em homogeneous} 
systems by Labb\'e \cite{L}, which we average over the
Brillouin zone determined by the external electric or magnetic modulation:
\begin{eqnarray}
\chi_1(q)=\frac{-1}{2\pi l^2}\sum_{n\sigma}\int_0^a\frac{dX_0}{a} 
\frac{\partial\nu_{n\sigma X_0}}{\partial\mu} 
\left[F_{nn}\left(\frac{(ql)^2}{2}\right)\right]^2 \,,
\label{chi1}
\end{eqnarray}
where we have used Eq.(\ref{flag}). 
We expect that this approximation is appropriate for weakly modulated 
systems characterized by $lV\ll a\hbar\omega_c$, or $lB_1\ll aB_0$.
Within the same procedure we evaluate $\chi_2$ as \cite{GG}
\begin{eqnarray}
\chi_2(q)=\frac{1}{2\pi l^2}\sum_{n\neq n'}
\left[F_{nn'}\left(\frac{(ql)^2}{2}\right)\right]^2 
\sum_{\sigma} \int_0^a\frac{dX_0}{a} 
\frac{\nu_{n\sigma X_0}-\nu_{n'\sigma X_0}}
{E_{n\sigma X_0}-E_{n'\sigma X_0}} \,.
\label{chi2}
\end{eqnarray}
For a sufficiently high magnetic field $B_0$, that is for sufficiently large 
energy gaps $E_{n+1,\sigma X_0}-E_{n\sigma X_0}$, and for a small $q$,
$\chi_2(q)$ is typically negligible with respect to $\chi_1(q)$, except 
eventually when there are very few states at the Fermi level such that 
$\chi_1(q)$ is vanishingly small.  Since for our weak modulations,
in the limit $q\rightarrow 0$, $\chi_1(q)$ becomes proportional to the
thermodynamic DOS, i.e. $\partial n/\partial \mu$, 
we can say the screening of the exchange interaction is dominated by the 
intra-Landau level transitions, which at low temperatures are determined 
by the DOS at the Fermi level.

This screened Hartree-Fock approximation (SHFA) has been used in the papers
by Ando, Ohkawa and Uemura for explaining the exchange enhancement of the 
Zeeman \cite{AU} and of the valley splitting \cite{OU} in homogeneous Si MOS 
systems. Further improvements have incorporated the dynamic screening, within 
the plasmon-pole approximation, and the Coulomb-hole correlation effects, 
for the calculation of the photoluminescence energy in n-doped GaAs quantum 
wells \cite{KA,US}.  Nevertheless, in order to keep the consistent
treatment of groundstate and transport properties tractable, we will 
neglect such additional corrections in the study of the modulated systems.  
In particular, even without dynamic screening effects, 
our results for the self-energy for a homogeneous 2DEG in a 
GaAs-AlGaAs interface are qualitatively similar to those shown in 
Refs. 33 and 34.
In Fig.2 we compare the first five spin splitted Landau levels vs. 
magnetic field, i. e. the Landau fan, in the standard (bare) HFA with the 
results given by the SHFA.  The material parameters are $m^*=0.067m_0$,
$g=0.4$, $\kappa=12.7$, and the carrier concentration is chosen such that
$\nu B_0=10$ T, $\nu$ being the filling factor. Both the Landau and the spin 
gaps are enhanced by the exchange interaction, but in the SHFA the Fock
self-energy is strongly dependent on the DOS at the Fermi level, $D(E_F)$.  
Thus the exchange interaction is
 screened for a high $D(E_F)$, i.e.
for noninteger filling factors, but it may become very large for an integer 
filling.  Since the exchange interaction is negative, the screening mechanism 
leads to the deep cusps in Fig.2(b), also present in the dynamic-screening 
calculations.\cite{KA,US}  Like in the HFA, the largest energy gaps occur 
for integer filling factors. 
The thermal energy is much smaller than the energy 
gaps, and no disorder broadening is assumed. Therefore the gaps at integer 
filling factors  in Fig.2(b) are smaller than those in Fig.2(a) only 
because of the screening involved by the response of the wave functions, 
described by $\chi_2$.

 We believe that the energy spectra, which we obtained within 
 the SHFA for the modulated system,
are much improved with respect to those in the HFA.  In Fig.3(a) 
we show the SHFA of the Landau bands generated by an electric  modulation 
of a period much larger than the magnetic length.  
The Zeeman splitting is enhanced in the states around the Fermi energy, with
replica in each of the upper and lower Landau bands, as in the HFA.\cite{MG}
The improvement consists in the recovering of the pinning effect,
resulting from  electrostatic screening,\cite{WGG} 
on the spin-splitted energy bands near the Fermi level.  
The pinning effect defines both
compressible (dispersionless) and spin polarized strips.  In the HFA, for the 
modulation parameters of Fig.3(a), strong short-range oscillations of the 
effective energies and of the charge density, with typical periods of
$3-5 l$, would occur.  Those oscillations would disappear only if the energy 
dispersion imposed by the external modulation would dominate the interaction 
effects, i.e. for a sufficiently short modulation period and/or for a 
sufficiently large modulation amplitude.\cite{MG} 
In sufficiently strong modulaton or steep confinement potentials the 
exchange enhancement of spin splitting is suppressed. With decreasing slope 
of the potential, the enhancement recovers suddenly, and a large spin 
splitting of the Landau bands occurs nearly symmetrically with respect to 
the Fermi energy in such a manner, that both spin polarized bands penetrate 
the Fermi energy with large slopes. Thus, the compressible strips obtained in 
the Hartree approximation are destroyed in the bare HFA and replaced with 
spin polarized incompressible strips. This spin polarization of the edge 
states may occur spontaneously even if the bare $g$-factor vanishes, and has 
been discussed as a type of phase transition.\cite{DGH,RB,SB} We believe that 
the occurrence of this spontaneous spin polarization is an artifact of the 
unscreened HFA, since our SHFA yields (as long as $a \gg l$) only a very 
smooth and gradual 
reduction of the spin splitting with increasing modulation strength, with 
compressible spin polarized strips (energies pinned to the Fermi energy) 
instead of incompressible ones.


In Fig.3(b) we transpose the results shown in Fig.3(a) for a magnetic
modulation.  In this case the Hartree (direct) screening is weaker, 
due to the weaker, purely quantum-mechanically, coupling of the charge 
density to the nonuniform magnetic field, and consequently the pinning
effect is less pronounced in Fig.3(b).  The peculiar Hartree response
to a nonuniform magnetic field, which becomes negligible in the classical 
limit (cyclotron radius at the Fermi energy larger than the modulation 
period) will be discussed in detail elsewhere.\cite{GMG}
In the present paper we restrict ourselves to a strong uniform component,
$B_0$, such that the electrostatic and the exchange effects due to the
periodic component are similar to those due to the electric modulation.

\section{Conductivities}

In order to calculate conductivities we have to consider the 
electron-impurity interaction.  For randomly distributed impurities, the 
self-energy $\Sigma^{ei}$ is diagonal with respect to $X_0$, and obviously 
also with respect to $\sigma$.  Since we are mostly interested in the 
effects due to the electron-electron interaction, we simplify the 
calculation of $\Sigma^{ei}$  by employing the phenomenological ansatz 
in which one ignores the dependence of its matrix elements on the Landau 
quantum number $n$ and on the center coordinate $X_0$, that is 
$\Sigma^{ei}_{n\sigma X_0,n'\sigma X_0}(E)\approx\Sigma^{ei}_{\sigma}(E)
\delta_{nn'}$.
In the SCBA, Fig.1, we need to solve the equation\cite{ZG}
\begin{eqnarray}
\Sigma^{ei}_{\sigma}(E)&=&\Gamma^2\sum_n
\int_0^a\frac{dX_0}{a}\,G_{n\sigma X_0}(E)\nonumber\\
&=&\Gamma^2\sum_n\int_0^a\frac{dX_0}{a}\,
\frac{1}{E-E_{n\sigma X_0}-\Sigma^{ei}_{\sigma}(E)}\,.
\label{scba}
\end{eqnarray}

We also take a simple parametrization of the electron-impurity interaction
energy, $\Gamma=\gamma\sqrt{B{\rm [Tesla]}}$ [meV].  
Using Eq.(\ref{scba}) the DOS can be written in the form
\begin{equation}
D(E)=\frac{\hbar\omega_c}{\pi\Gamma^2}D_0
\sum_{\sigma}Im\,\Sigma^{ei}_{\sigma}(E)\,,
\label{dos}
\end{equation}
where $D_0=m^*/2\pi\hbar^2$ is the DOS per spin level for the homogeneous
system without magnetic field.

The standard Kubo formula for the conductivity tensor can be written as
\begin{equation}
\sigma_{\alpha\beta}(\omega)=e\int_0^{\infty}dt
\frac{e^{i\omega t}-1}{\hbar\omega}e^{-\varepsilon t}
Tr\langle{\cal F}(H)[\dot r_{\beta},\,
e^{\frac{i}{\hbar}Ht}j_{\alpha}e^{-\frac{i}{\hbar}Ht}]\rangle_{imp} \,,
\label{kubo1}
\end{equation}
where $\alpha,\beta=x,y$, $r_{\alpha}\equiv \alpha$, and 
$\varepsilon\rightarrow 0^+$.  The operator $\dot {\bf r}$ is given by
\begin{equation}
\dot {\bf r}=\frac{i}{\hbar}[H,{\bf r}]\equiv 
{\bf v}^0+{\bf v}^{ex}\,,
\label{xdot}
\end{equation}
where ${\bf v}^0$ is the usual form of the velocity operator, resulting from 
the commutator of the noninteracting Hamiltonian $H^0$ with the position,
\begin{equation}
{\bf v}^0=\frac{1}{m^*}\left({\bf p}+e{\bf A(r)}\right)\,.
\label{v0}
\end{equation}
All the other terms of the full Hamiltonian $H$ commute with ${\bf r}$
except the Fock self-energy, due to its nonlocal character. 
According to Eqs.(\ref{hfa}) and (\ref{parden}), the Coulomb interaction.  
gives an exchange contribution to the velocity operator, of the form
\begin{equation}
{\bf v}^{ex}=\frac{i}{\hbar}[\Sigma^{ee,F},{\bf r}]\,.
\label{vex}
\end{equation}
The matrix elements of the velocity operators 
(\ref{v0}) and (\ref{vex}), in the Landau basis, can be found in Appendix B.
For the current-density operator we adopt the definition  
\begin{equation}
{\bf j}=-\frac{e}{L_x L_y} \dot {\bf r}\,,
\end{equation}
and we will call the term corresponding to Eq.(\ref{vex}) "exchange current".
After a straightforward manipulation the Kubo formula can be
put in the form
\begin{equation}
\sigma_{\alpha\beta}(\omega)=\frac{1}{i\omega}
[\lambda_{\alpha\beta}(\omega)-\lambda_{\alpha\beta}(0)]\,,
\label{kubo2}
\end{equation}
with
\begin{eqnarray}
&&\lambda_{\alpha\beta}(\omega)=\frac{-e^2}{L_xL_y}\int_{-\infty}^{\infty} dE 
{\cal F}(E)\nonumber\\
&&\times\,\,
Tr\left\{\langle\delta(E-H)\dot r_{\alpha}\widehat G^+(E+\hbar\omega)
\dot r_{\beta}\rangle_{imp}+
\langle\dot r_{\beta}\widehat G^-(E-\hbar\omega)\dot r_{\alpha}\delta(E-H)
\rangle_{imp}\right\}\,.
\label{lambda}
\end{eqnarray}

Of course $\dot{\bf r}$ depends on the impurity configuration, via the 
exchange contribution.  
However, since in the following we will restrict ourselves to a weak disorder, 
the dependence of Eq.(\ref{vex}) on the disorder broadening will be 
negligible.  Therefore we neglect the impurity effects on 
$\dot {\bf r}$ in Eq.(\ref{lambda}). Then, making use of the formal relation 
$\delta(E-H)=[\widehat G^-(E)-\widehat G^+(E)]/2\pi i$, 
we find that in the static limit, $\omega\rightarrow 0$, we have to 
average combinations like
$\langle \widehat G^{\pm}(E)\dot r_{\alpha}\widehat G^{\pm}(E)\rangle_{imp}$.
The result can be written as 
\begin{eqnarray}
&&\langle\widehat G^{\pm}(E)\dot r_{\alpha}\widehat G^{\pm}(E)\rangle_{imp}
\nonumber\\
&&=-\frac{e}{L_xL_y}\langle \widehat G^{\pm}(E) \rangle_{imp}
\left(\dot r_{\alpha}+\frac{i}{\hbar}[\Sigma^{ei}(E),r_{\alpha}]\right)
\langle \widehat G^{\pm}(E) \rangle_{imp} \,,
\label{vcor}
\end{eqnarray}
where we have used a Ward identity to express the current vertex correction
due to the impurities with the help of the corresponding self-energy 
operator.\cite{G2}  Since we have assumed $\Sigma^{ei}(E)$ 
to be a trivial 
(spin-dependent) c-number, the impurity vertex correction vanishes.
But we see that the Coulombian vertex correction is automatically included
in Eq.(\ref{vcor}) via the exchange current. 

The Hall conductivity can now be written as\cite{ZG}
\begin{eqnarray}
&&\sigma_{xy}=-\sigma_{yx}=\frac{\hbar e^2}{i\pi^2l^2}
\int_{-\infty}^{\infty}dE{\cal F}(E)\nonumber\\
&&\times\int_0^a\frac{dX_0}{a}\sum_{nn'\sigma}
\langle\psi_{n\sigma X_0}\mid\dot x\mid\psi_{n'\sigma X_0}\rangle
\langle\psi_{n'\sigma X_0}\mid\dot y\mid\psi_{n\sigma X_0}\rangle\nonumber\\
&&\times Im\, G_{n\sigma X_0}(E)\,Re\,\frac{d}{dE} G_{n'\sigma X_0}(E)\,.
\label{chall}
\end{eqnarray}
Since we treat the disorder in the SCBA, we cannot consider localization 
effects, and hence we cannot expect real Hall plateaus.  Moreover, if the 
disorder broadening of the Landau levels is much smaller than the energy 
bands, Eq.(\ref{chall}) can be well approximated by the limit of a vanishing 
electron-impurity interaction, $\gamma\rightarrow 0$.  In this limit 
$\sigma_{xy}$ takes the more familiar form 
\begin{eqnarray}
&&\sigma_{xy}=\frac{i\hbar e^2}{\pi l^2}\int_0^a\frac{dX_0}{a}
\sum_{n\neq n',\,\sigma}{\cal F}(E_{n\sigma X_0})\nonumber\\
&&\times\frac
{\langle\psi_{n\sigma X_0}\mid\dot x\mid\psi_{n'\sigma X_0}\rangle
\langle\psi_{n'\sigma X_0}\mid\dot y\mid\psi_{n\sigma X_0}\rangle}
{(E_{n\sigma X_0}-E_{n'\sigma X_0})^2}\,,
\label{chall0}
\end{eqnarray}
in which the energy gaps are explicitly evidenced.

It is instructive to mention that neglecting the exchange current 
one obtains deviations from the well known quantized values, $\sigma_{xy}=
(e^2/h)\times(integer)$, for an integer filling factor. Let us
consider the simplest case, with no external modulation. In the absence 
of the Coulomb interaction the energy gaps are determined by the cyclotron 
energy. The cyclotron frequency squared, in the denominator of 
Eq.(\ref{chall0}), is compensated by the cyclotron frequencies introduced in
the numerator by the velocity matrix elements, Eq.(\ref{vit0}), 
and for a sufficiently low temperature one obtains
the simple Drude formula, $\sigma_{xy}=(e^2/h)\nu$.   When the Coulomb
interaction is present, the energy gaps are enhanced, but a similar 
enhancement occurs in the numerator, determined by the exchange term of
the velocities, Eqs.(\ref{vex}) and (\ref{vitex}), and the Drude formula 
remains valid.  Obviously, when the exchange interaction  is screened
the exchange currents may be small.   In our calculations, for 
the modulated systems, their contribution to the conductivities have been of 
the order 30-50~\% in the HFA,\cite{MGT} but of no more than 10 \% in the 
present SHFA.

The longitudinal conductivities 
can be transformed from Eqs.(\ref{kubo2})-(\ref{lambda}) to
\begin{eqnarray}
&&\sigma_{\alpha \alpha}=\int dE \left( -\frac{d{\cal F}}{dE} \right)
\sigma_{\alpha \alpha}(E)\,,\nonumber\\
&&\sigma_{\alpha \alpha}(E)=\frac{\hbar e^2}{l^2\pi^2}\int_0^{a} 
\frac{dX_0}{a} 
\sum_{nn'\sigma}\mid\langle \psi_{n\sigma X_0}\mid 
\dot r_{\alpha}\mid \psi_{n'\sigma X_0}\rangle\mid ^2\nonumber\\
&&\times\,Im \, G_{n\sigma X_0}(E) \, Im \, G_{n'\sigma X_0}(E) \,,
\label{clong}
\end{eqnarray}

The relation between the longitudinal conductivities and the DOS 
is complicated. Both the diagonal conductivities $\sigma_{xx}$ 
and $\sigma_{yy}$ have inter-level components, corresponding 
to $n\neq n'$ in Eq.(\ref{clong}), also known as {\it scattering}
conductivities.  
%
%
One can show, by inspecting Eq.(\ref{clong}), that for a weak disorder,
$\gamma\rightarrow 0$, the scattering conductivities become proportional to 
$(\Gamma D(E_F))^2$. Hence, they qualitatively  reproduce the DOS 
profile, with van Hove singularities corresponding to the edges of the 
one-dimensional Landau bands.

Due to the anisotropy  of the system, an intra-level component of 
Eq.(\ref{clong}), with $n=n'$, is nonzero for $\sigma_{yy}$ only. 
It is called {\it band} conductivity,\cite{AV,ZG}
being directly related to the dispersion of the Landau bands which
yields equilibrium Hall currents.  If the Hamiltonian is a local
operator, the Hellman-Feynman theorem leads to
\begin{equation}
\frac{d E_{n\sigma X_0}}{dX_0}=-m\omega_c
\langle\psi_{n\sigma X_0}\mid\dot y\mid\psi_{n\sigma X_0}\rangle\,,
\label{feyhel}
\end{equation}
which is no longer valid when the exchange interaction is considered. 
However, if the latter is screened, the deviations from Eq.(\ref{feyhel}) 
are not very important, and  the band conductivity can 
still be understood in terms of the energy dispersion.  Classically, it 
corresponds to the quasi-free drift of the centers of the cyclotron 
orbits along the $y$ axis, parallel to the periodic electric field.
Thus the band conductivity behaves contrary to the scattering components:
It vanishes at the band edges, has maxima at the band centers, and 
{\it diverges} for a small disorder, like $\gamma^{-2}$, 
becoming the dominant contribution to $\sigma_{yy}$.                      

The structure of the longitudinal 
conductivities, $\sigma_{xx}(E)=\sigma_{xx}^{scattering}(E)$ and 
$\sigma_{yy}(E)=\sigma_{yy}^{scattering}(E)+\sigma_{yy}^{band}(E)$, for an 
isolated, sinusoidal Landau band are qualitatively displayed in Fig.4.  
In our calculations the two scattering conductivities are nearly equal.  
In Fig.4(a) we assume a small band conductivity, such that both $\sigma_{xx}$
and $\sigma_{yy}$ show the two van Hove peaks.  Reducing the disorder
broadening the scattering components decrease, but at the same time 
the band conductivity increases, and a triple-peak structure evolves in 
$\sigma_{yy}$, Fig.4(b), while the shape of $\sigma_{xx}$ remains unchanged.  
For an even smaller disorder $\sigma_{yy}$ becomes dominated by the band 
conductivity, the central peak covering the vHS, like in Fig.4(c).

\section{Discussion of the numerical results}

We perform the calculations of the effective electronic states within a 
numerical iterative scheme, starting from the noninteracting solution,
and assuming a fixed number of particles for determining the chemical
potential.  At each step we diagonalize the Hamiltonian matrix, 
Eqs.(\ref{h0fs}), (\ref{htfs}) and (\ref{ftfs}), in $L$ points of the half
Brillouin zone, $0\leq X_0\leq {\pi}/K$, and then we compute the first $L$ 
Fourier harmonics for Eqs.(\ref{wffs}) and (\ref{enfs}).  We have taken $L$ 
in the range 10-40, and we have mixed 5-10 Landau levels.  Then, we use 
Eqs. (\ref{chall0}) and (\ref{clong}) to calculate the conductivity tensor.  

In order to resolve both the van Hove peaks in the modulated system
and the spin splitting of the Landau bands, we need a very small disorder 
parameter.   In Fig.5 we show the magnetoconductivity tensor for a magnetic 
field $B_0$ varied such that the Fermi level traverses the Landau bands 
with $n=1$ and $n=2$. In Fig.5(a) we consider a pure electric modulation
of amplitude $V=15$ meV and period $a=500$ nm.  For a better understanding
of the conductivity oscillations, three qualitatively different energy 
band structures are indicated in Fig.6, in a half Brillouin zone.  The 
bands corresponding to opposite spin directions are separated for filling 
factors below 4, i.e. for $B_0>2.5$ T, due to the exchange enhancement, 
and they partially overlap for higher filling factors.

For the electric modulation the energy dispersion is strongly dependent 
on $D(E_F)$, due to the strong screening contained both in $\Sigma^{ee,H}$
and in $\Sigma^{ee,F}$.  All the Landau bands shrink whenever a band is 
intersected by the Fermi level.  Therefore a full spin splitting --- i.e. 
non-overlapping bands with the same $n$, but with different 
spin directions --- can result even for a modulation amplitude much 
larger than $\hbar\omega_c$.

Due to the small band width, and also due to the pinning of the band edges to 
the Fermi level, the resolution of the vHS peaks requires a very small 
disorder broadening. For our parameters the vHS can be distinguished in 
$\sigma_{xx}$, Fig.5, and their resolution even improves with decreasing 
magnetic field.  The reason is the 
increase of the exchange interaction, self-consistently with the reduction
of the DOS: $D(E_F)$ decreases, the screening diminishes, and the exchange 
broadening is larger.  Nevertheless, the screening may again increase for
overlapping bands, like in Fig.6(c).  In that case two bands contribute
to the pinning effect, and therefore the resolution of the vHS in the states 
$(2,+,0)$ and $(2,-,\pi/K)$, which yield the two maxima of $\sigma_{xx}$
in Fig.5(a) for 1.9 T $<B_0<$ 2.1 T, is poorer than in the states
$(2,+,\pi/K)$ and $(2,-,0)$, that is for 2.4 T $<B_0<$ 2.5 T and
1.6 T $<B_0<$ 1.7 T, respectively.

While the scattering component of $\sigma_{yy}$ is nearly identical to 
$\sigma_{xx}$, the very small disorder parameter makes the band conductivity 
large. In Fig.5(a) $\sigma_{yy}$ is thus in the situation depicted in 
Fig.4(c), with the vHS profile hidden by the band-conductivity single peak.  
Decreasing the magnetic field, due to the stronger energy dispersion the band 
conductivity further increases,  contrary to the scattering term, which 
becomes negligible. For the overlapping bands of Fig.6(c) the minimum of 
$\sigma_{yy}$ at $B_0=2$ T corresponds to the strongest screening, i.e. to 
the highest $D(E_F)$. 

In Fig.5(b) we show similar results obtained for a purely magnetic modulation 
which produces Landau bands equivalent to those of Fig.6.  We have chosen  
$B_1=0.25$ T, the other parameters of the calculation being the same as those
for Fig.5(a). There are two, but unessential differences with respect to the  
electric modulation of Fig.5(a). First, the energy dispersion increases with 
increasing Landau quantum number $n$. In a crude approximation, i.e. 
neglecting the Coulombian effects, as long as $Kl\ll 1$ the energy bands are 
determined by the local cyclotron energy, 
$E_{n\sigma X_0}=\frac{\hbar e}{m^*}(B_0+B_1\cos KX_0)(n+\frac{1}{2})$.  
Second, as mentioned at the end of Sec.II, the Hartree (direct) screening is 
weaker than in the electric case, and hence the strength of the exchange with 
respect to the Hartree interaction may increase
to some extent.  Going through Fig.5(b) from high to low values of the 
constant field $B_0$, the increase of the energy dispersion at the Fermi 
level due to the increase of the quantum number $n$ is thus amplified by the 
exchange broadening.  For filling factors higher than 4, four Landau bands 
overlap, i. e. those with $n=2$ and $n=3$.  The order of the vHS observed in 
$\sigma_{xx}$, for 
$B_0<2.5$ T is : $(2,+,\pi/K)$ ($B_0=2.48$ T), $(2,-,\pi/K)$ ($B_0=2.10$ T), 
$(2,+,0)$ ($B_0=1.84$ T), $(3,+,\pi/K)$ ($B_0=1.72$ T).  The small scattering 
conductivity $\sigma_{xx}$ for 2.15~T~$<B<$~2.40~T also reflects a strong 
energy dispersion.  Additionally, the screening-exchange balance may  
generate weak DOS fluctuations near the band edges,\cite{MGT} which can be 
seen as the shoulders of the band conductivity  (via the Green function 
squared in the Kubo formula), in $\sigma_{yy}$ for 2.0~T~$<B_0<$~2.5~T.  
For lower magnetic fields, when $D(E_F)$ increases 
due to the overlapping bands, the exchange effects are again small, and 
the maxima of the scattering conductivity corresponds to the minima
of the band conductivity and vice-versa.

For comparison with experiment we need to invert the conductivity
tensor into resistivities,
\begin{equation}
\rho_{xx}=\frac{\sigma_{yy}}{{\cal D}}\,,\,
\rho_{yy}=\frac{\sigma_{xx}}{{\cal D}}\,,\,
\rho_{xy}=\frac{\sigma_{yx}}{{\cal D}}\,,
\end{equation}
where ${\cal D}=\sigma_{xx}\sigma_{yy}+\sigma_{xy}^2$. In our calculations  
$\sigma_{xx}\sigma_{yy}\ll\sigma_{xy}^2$, and hence we have 
$\rho_{xx,yy}\approx\sigma_{yy,xx}/\sigma_{xy}^2$.  In Fig.5(c) we see
that the resistivity measured perpendicular to the modulation
reflects in fact the band conductivity, while the scattering
conductivity alone can be observed only in the resistivity measured
parallel to the modulation.

In the experiments on the systems with an etched electric modulation, of
parameters comparable to our model, more complicated structures of 
the resistivity $\rho_{xx}$ have been observed.\cite{MGT}  The 
SdH maxima corresponding to a certain spin direction may show
two and even three internal peaks.  One possible explanation could 
consist in the more complicated exchange effects on the band conductivity,
like the shoulders we have found for the magnetic modulation, Fig.5(b).
We have also obtained such effects for an electric modulation, with a shorter 
period, $a=100$ nm, when the screening is considerably reduced.\cite{MGT}  
Another possible explanation, which is supported by the present results
of the SHFA, is  that in the real system the band and the scattering
conductivities may be comparable in magnitude, such that their superposition
in $\sigma_{yy}$ may lead to one to three peaks per Landau level, as
illustrated in Fig.4.  However, within our approximations, the price for
the resolution of both the spin splitting and the vHS is a relatively large
band conductivity.

To our knowledge, a simultaneous measurement of both $\rho_{xx}$ and 
$\rho_{yy}$, capable of indicating the real magnitude of the two 
components of the longitudinal conductivities, has been reported, at high 
magnetic fields, only for the short-period modulated systems,\cite{HPE}
and $\rho_{xx}$ has been found considerably larger than $\rho_{yy}$, but
without a clear DOS structure.  For short periods, the Hartree screening 
gets weaker and the Landau bands become wider, such that in principle we 
could cover several situations, including those of Fig.4, by varying the
modulation amplitude and the disorder parameter $\gamma$. We cannot extend 
our SHFA to that regime, because of the steepness of the energy dispersion 
which is not compatible with the assumption about the quasi-homogeneous 
screening of the exchange interaction. Nevertheless, the results of the HFA 
may be satisfactory.\cite{MGT}

We want now to discuss the situation of an electric modulation with a larger 
amplitude, such that the Landau bands overlap.  In the experiment by Weiss 
et. al. \cite{WKP2} the double-peak structure of the resistivity $\rho_{xx}$ 
has been found, while the absence of the spin splitting for the unmodulated
system may be attributed to the combined disorder and thermal effects. We 
have recently confirmed such a possibility, within the SHFA, by choosing such 
a disorder parameter and temperature that both the exchange enhancement of 
the Zeeman splitting and the band conductivity have been suppressed,
and $\rho_{xx}$ have been obtained as resulting from Fig.4(a).\cite{MGW} 

In the example of Fig.7 we consider the situation in which neither the spin 
splitting nor the vHS are resolved in the low-energy 
Landau bands, but the vHS develop in the higher, overlapped bands. 
The SdH minima at filling factors around 2 and 4 are slightly shifted
to higher magnetic fields.  The energy gaps are small, nearly vanishing,
but the adjacent bands are separated due to the pinning effect. For weaker 
magnetic fields, at even filling factors the vHS from the top of the Landau 
band below the Fermi level overlaps with that from the bottom of the upper 
band, resulting in the maxima of $\sigma_{xx}$ for $\nu=8$ and $\nu=10$
and in the minima for the odd filling factors $\nu=7$ and $\nu=9$.  The 
transition occurs around $\nu=6$. Since the DOS decreases when the magnetic 
field is lowered, the screening becomes less efficient and the band width may 
increase when the Fermi level is in a band center, leading to a large
band conductivity, wherefrom the maxima of $\sigma_{yy}$ for $\nu=7$ and 
$\nu=9$.  The switching from even to odd filling factors at the resistivity 
minima, which we show here only for $\rho_{yy}$ ($\sigma_{xx}$), has been 
recently observed for $\rho_{xx}$ and supported by a transport calculation
similar to the present one, but with the Coulomb interaction
neglected.\cite{TWM}

Finally we want to mention that the plateaus we obtain for the Hall 
resistivity, Fig.7(b), do not correspond to plateaus of the Hall conductivity,
since localization effects do not exist in our model, but to the wide minima
of the longitudinal conductivities.  In the noninteracting approximation, 
and also in the HFA, for a fixed number of particles, the Fermi level jumps 
abruptly from one Landau band to the other.  In the SHFA the exchange 
interaction is very sensitive to the variation of $D(E_F)$, and since they 
are self-consistently determined, the stability of the chemical potential 
in an energy gap considerably improves.

\section{Conclusions and final remarks}

We have calculated energy spectra of a two-dimensional electron gas in an 
electric or a magnetic superlattice, and in a perpendicular magnetic field, 
beyond the standard HFA.  The superlattice potential is smooth and the uniform 
magnetic field is strong, such that the electrostatic screening is very 
important.  The essential element of our approach is the inclusion of
screening in the exchange term, self-consistently with the DOS at the Fermi 
level. The numerical calculations are complicated and time consuming, such 
that approximations in treating the screening are unavoidable: we have only 
considered the static screening, in a manner appropriate to a 
quasi-homogeneous system. Improvements of our procedure are possible, e.g. 
along the lines of 
Refs.~32,~33,
but even within the present approach the results drastically change 
with respect to the HFA.  
%
The obtained screening of the exchange interaction may be even somewhat too 
strong, since we found that our SHFA yields no charge-density-wave 
instability of the homogeneous 2DEG, even at very low filling factors.

Using a large-amplitude modulation as  model 
for edge states, we have shown that our SHFA
results interpolate
between the contradictory Hartree and Hartree-Fock approximations:  we
obtained both compressible edge strips much wider than the magnetic length,
and an enhanced spin splitting. To the best of our knowledge this expected 
result has never been obtained within a microscopic many-body theory before.

For a weak modulation, the standard HFA yields strong short-range oscillations
of the Landau bands and the particle density, which have never been observed
experimentally.  Such oscillations completely disappear in our SHFA, and we 
believe that these present results are much more realistic.
Clearly, we cannot expect a quantitative agreement with experiments since, 
besides the approximations regarding the
electron-electron interaction, we have used a rather crude simplification
of the electron-impurity scattering in our transport calculation, such that
the disorder vertex correction vanishes.  But qualitatively, we have
obtained smooth internal structures of the SdH peaks determined by the
interplay of the scattering and band conductivities, the ingredients of 
which have been carefully analyzed. These smooth internal structures 
compare much more favorably with the experimental results than the sharp 
structures near the Landau band edges obtained in the bare HFA.\cite{MGT}


\acknowledgments{
We wish to thank Marc Tornow, Dieter Weiss, Behnam Farid, and Paul Gartner 
for discussions.  One of us (A.M.) is also grateful to the 
Max-Planck-Institut f\"ur Festk\"orperforschung, Stuttgart, for support 
and hospitality.}

\appendix

\section{Hamiltonian matrix elements}

With the notations $z=(Kl)^2/2$ and
\begin{equation}
F_{nn'}(z)=\left(\frac{n'!}{n!}\right)^{1/2}z^{(n-n')/2}
e^{-z/2}L_{n'}^{n-n'}(z),
\label{flag}
\end{equation}
where $L_{n'}^{n-n'}(z)$ are the Laguerre polynomials,\cite{AS} 
$n,n'=0,1,...$, and $F_{nn'}=(-1)^{n-n'}F_{n'n}$, we can write the matrix 
elements of the noninteracting Hamiltonian in the form
\begin{eqnarray}
&&H^0_{nn'}(\sigma,X_0)=\delta_{nn'}\hbar\omega_c
\Bigg[n+\frac{1}{2}+\frac{1}{8z}\left(\frac{B_1}{B_0}\right)^2\Bigg]
-\delta_{nn'}\frac{\sigma}{2}g\mu_BB_0\nonumber\\
&&+\Bigg[\frac{\hbar\omega_c}{2z}\frac{B_1}{B_0}
\left(F_{nn'}(z)+\sqrt{nn'}F_{n-1,n'-1}(z)-
\sqrt{(n+1)(n'+1)}F_{n+1,n'+1}(z)\right)\nonumber\\
&&+\left(V-\frac{\sigma}{2}g\mu_BB_1\right)F_{nn'}(z)\Bigg]
\cos\left(KX_0+(n-n')\frac{\pi}{2}\right)\nonumber\\
&&-\frac{\hbar\omega_c}{8z}\left(\frac{B_1}{B_0}\right)^2
F_{nn'}(4z)\cos\left(2KX_0+(n-n')\frac{\pi}{2}\right)
\label{h0fs}
\end{eqnarray}
Due to the reflection symmetry of the external fields, we have the following 
parity rule,
\begin{equation}
H^0_{nn'}(\sigma,X_0)=(-1)^{n-n'}H^0_{nn'}(\sigma,-X_0)\,,
\label{parul}
\end{equation}
which also holds for the Coulomb self-energy matrix elements, Eq.(\ref{hfa}),
which depend on the self-consistent wave functions given by Eq.(\ref{effwf}). 
In order to write those matrix elements in a form suitable for a numerical 
calculation, we expand the mixing coefficients in Fourier series,
\begin{equation}
c_{nn'}(\sigma,X_0)=\sum_{p\geq 0}\gamma_{nn'}(\sigma,p)
\cos\left(pKX_0+(n-n')\frac{\pi}{2}\right)\,,
\label{wffs}
\end{equation}
where we have taken into account the parity rule.

The effective energies can thus be expanded as
\begin{equation}
E_{n\sigma X_0}=\sum_{p\geq 0} \epsilon_{n\sigma p} \cos pKX_0\,,
\label{enfs}
\end{equation}
and similarly the occupation numbers
\begin{equation}
\nu_{n\sigma X_0}=\sum_{p\geq 0} f_{n\sigma p} \cos pKX_0\,.
\label{fffs}
\end{equation}

We define 
\begin{equation}
A_{p_1p_2p_3p_4}^{n_1n_2n_3n_4}=\frac{4}{\pi}\int_{-\pi}^{\pi}
\,dx\,\prod_{i=1}^4\cos(p_ix+n_i\frac{\pi}{2}).
\end{equation}
With this notation, and using Eqs.(\ref{wffs}) and (\ref{fffs}), the Fourier 
coefficients of the particle density can be written, by inverting
Eq.(\ref{parden}), as
\begin{eqnarray}
&&n_p=\frac{2-\delta_{p0}}{16\pi l^2}
\sum_{^{m_1\geq 0,\sigma=\pm}_{p_1\geq 0}}
f_{m_1\sigma p_1}\nonumber\\
&&\times\sum_{^{m2,m3\geq 0}_{p2,p3\geq 0}}
A_{p~~~~p_1~~~~p_2~~~~p_3}^{m_2-m_3,0,m_1-m_2,m_1-m_3}
F_{m_2m_3}(p^2z)\gamma_{m_1m_2}(\sigma,p_2)\gamma_{m_1m_3}(\sigma,p_3).
\label{defs}
\end{eqnarray}

The matrix elements of the Hartree term, defined by Eqs.(\ref{hfa})
and (\ref{parden}), can now be put in the form
\begin{equation}
\Sigma_{nn'}^{ee,H}(X_0)=\frac{2\pi e^2}{\kappa Kl}
\sum_{p\geq 1}\frac{n_p}{p}F_{nn'}(p^2z)\cos\left(pKX_0+(n-n')
\frac{\pi}{2}\right)\,,
\label{htfs}
\end{equation}
where we have used the Fourier transform of the Coulomb potential.
The corresponding Fourier series for the Fock term of the Hamiltonian,
Eq.(\ref{fter}), can be found by directly searching for the Fourier
amplitudes, and after a lengthy, but straight forward calculation, one
gets
\begin{eqnarray}
&&\Sigma_{nn'}^{ee,F}(\sigma,X_0)=-\frac{1}{16\pi^2 l}
\sum_{p\geq 0}(2-\delta_{p0})\cos\left(pKX_0+(n-n')\frac{\pi}{2}\right)
\sum_{^{m_1\geq 0}_{p_1\geq 0}}f_{m_1\sigma p_1}\nonumber\\
&&\times\sum_{^{m2,m3\geq 0}_{p2,p3\geq 0}}
A_{p~~~~p_1~~~~p_2~~~~p_3}^{m_2-m_3,0,m_1-m_2,m_1-m_3}
S_{nm_2,n'm_3}(pKl)\gamma_{m_1m_2}(\sigma,p_2)\gamma_{m_1m_3}(\sigma,p_3)\,,
\label{ftfs}
\end{eqnarray}
where S denote the exchange integrals 
\begin{eqnarray}
&&S_{m_1n_1,m_2n_2}(t)=\left(\frac{m_1!m_2!}{n_1!n_2!}\right)^{1/2}
\int_0^{\infty}\,dq\,\tilde u(q\sqrt{2})\nonumber\\
&&\times e^{-q^2}q^{n_1-m_1+n_2-m_2+1}
J_{n_1-m_1-n_2+m_2}(tq\sqrt{2})L_{m_1}^{n_1-m_1}(q^2)
L_{m_2}^{n_2-m_2}(q^2)\,,
\end{eqnarray}
with $J_n(x)$ the Bessel functions.
For the homogeneous system the self-consistent wave functions
are identical with the Landau wave functions, so that
$\gamma_{n_1n_2}(\sigma,p)=\delta_{n_1n_2}\delta_{p0}$, and the
exchange integrals reduce to $S_{m_1n_1,m_1n_1}(0)$. \cite{MG,DOL}

\section{Velocity matrix elements}

The matrix elements of the noninteracting components of the velocity 
operators, Eq.(\ref{v0}), in the absence of a magnetic modulation, are
\begin{equation}
(v^0_{\alpha})_{nn'}(X_0)=\eta_{\alpha}\frac{l\omega_c}{\sqrt{2}}
\left(\sqrt{n+1}\,\delta_{n',n+1}+
\eta_{\alpha}^2\sqrt{n'+1}\,\delta_{n,n'+1}\right)
\label{vit0}
\end{equation}
with the notation $(\eta_x,\eta_y)=(i,1)$.  When the magnetic field has a 
periodic component in the $x$ direction, one has to add to $v^0_y$ the 
extra term
\begin{equation}
\frac{\omega_cB_1}{KB_0}F_{nn'}(z)
\sin\left(KX_0+(n-n')\frac{\pi}{2}\right)\,.
\end{equation}

The matrix elements of the exchange current can be transformed in Fourier 
series like those of the exchange-interaction operator, as sketched in 
Appendix A.  According to Eq.(\ref{vex}), the form of 
${\bf v}^{ex}_{nn'}(\sigma,X_0)$ is similar to that of 
$\Sigma_{nn'}^{ee,F}(\sigma,X_0)$, with the replacement
$u({\bf r}-{\bf r}')\rightarrow({\bf r}-{\bf r}')u({\bf r}-{\bf r}')$
in Eq.(\ref{hfa}), and, after an integration by parts, with
$\tilde u({\bf q})\rightarrow\nabla_{\bf q}\tilde u({\bf q})$ in 
Eq.(\ref{fter}), one obtains the following Fourier expansion: 
\begin{eqnarray}
&&{\bf v}^{ex}_{nn'}(\sigma,X_0)=\frac{1}{8\sqrt{2}\pi h}
\sum_{p\geq 0}(2-\delta_{p0})\sin\left(pKX_0+(n-n')\frac{\pi}{2}\right)
\sum_{^{m_1\geq 0}_{p_1\geq 0}}f_{m_1\sigma p_1}\nonumber\\
&&\times\sum_{^{m2,m3\geq 0}_{p2,p3\geq 0}}
A_{p~~~~p_1~~~~p_2~~~~p_3}^{m_2-m_3,0,m_1-m_2,m_1-m_3}{\bf T}_{nm_2,n'm_3}
(pKl)\gamma_{m_1m_2}(\sigma,p_2)\gamma_{m_1m_3}(\sigma,p_3)\,,
\label{vitex}
\end{eqnarray}
in which  
\begin{eqnarray}
&&(T_{\alpha})_{m_1n_1,m_2n_2}(t)=\eta_{\alpha}
\left(\frac{m_1!m_2!}{n_1!n_2!}\right)^{1/2}
\int_0^{\infty}\,dq\,\left[\frac{d\tilde u(q\sqrt{2})}{dq}\right]
\nonumber\\
&&\times e^{-q^2}q^{n_1-m_1+n_2-m_2+1}
[J_{n_1-m_1-n_2+m_2-1}(tq\sqrt{2})-\eta_{\alpha}^2
J_{n_1-m_1-n_2+m_2+1}(tq\sqrt{2})] \nonumber\\
&&\times L_{m_1}^{n_1-m_1}(q^2)
L_{m_2}^{n_2-m_2}(q^2)\,.
\end{eqnarray}

\begin{figure}
\caption{Diagrammatic representation of the combined screened Hartree-Fock 
and self-consistent Born approximations.  The wavy lines are the Coulomb
interaction and the dashed line is the electron-impurity interaction.}
\end{figure}

\begin{figure}
\caption{The Landau fan (a) in the HFA and (b) in the SHFA for a homogeneous
GaAs system.  The dashed line shows the chemical potential.
The temperature $T=1$ K.}
\end{figure}

\begin{figure}
\caption{Compressible and incompressible strips in the SHFA, for: (a) an 
electric modulation having $a=1000$ nm and $V=200$ meV with $B_0=6$ T, and
(b) a magnetic modulation having $a=500$ nm and $B_1=1.2$ T and $B_0=4$ T.
The Fermi level is indicated by the horizontal dashed line. $T=1$ K.}
\end{figure}

\begin{figure}
\caption{Three possible profiles of the magnetoconductivity corresponding 
to a sinusoidal-like isolated Landau band.  The full line shows $\sigma_{yy}$, 
the dashed lines $\sigma_{xx}$ and the scattering component of $\sigma_{yy}$,
and the dotted lines the band conductivity.} 
\end{figure}

\begin{figure}
\caption{The conductivity tensor calculated for (a) an electric modulation
with $a=500$ nm and $V=15$ meV, (b) a magnetic modulation with
$a=500$ nm and $B_1=0.25$ T, and (c) the resistivities corresponding
to (a). The $xx$ components are magnified by a factor of two.  
The temperature $T=1$ K and the disorder parameter $\gamma=0.06$.}
\end{figure}

\begin{figure}
\caption{Three energy spectra corresponding to Fig.5a.}
\end{figure}

\begin{figure}
\caption{(a) The conductivities, and (b) the resistivities, for an electric
modulation with $a=500$ nm and $V=20$ meV.  The Hall components are
reduced by a factor of two.  The numbers inside the plot indicate the filling 
factors.  $T=1$ K and $\gamma=0.20$.}
\end{figure}

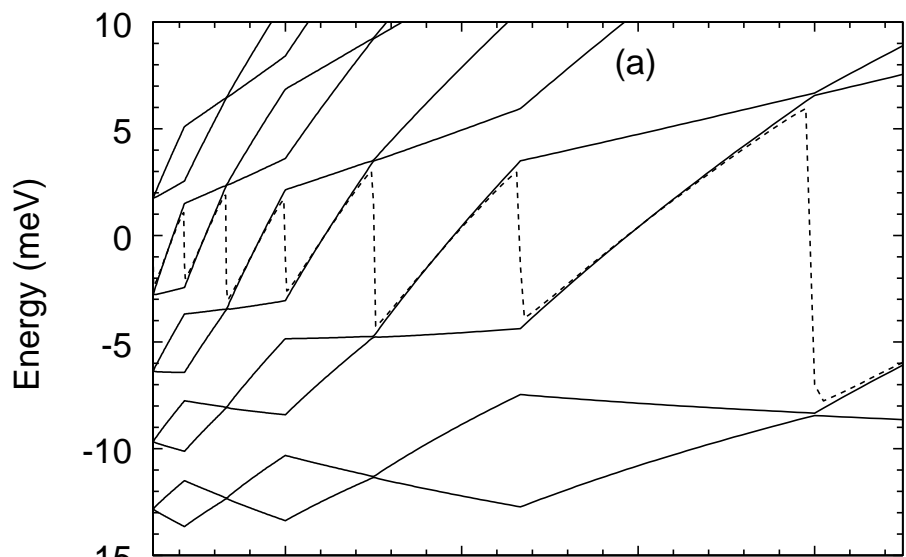
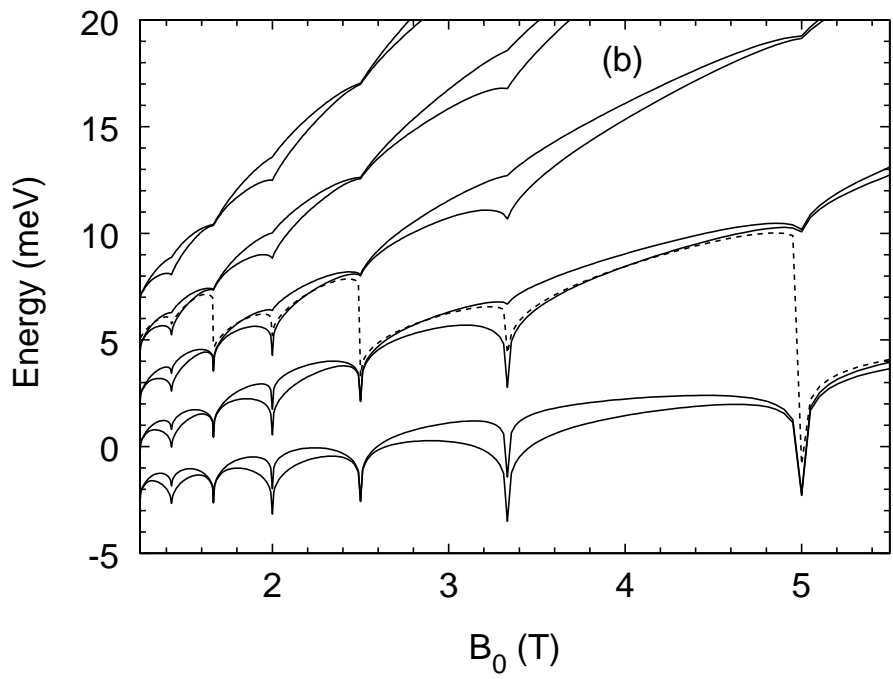
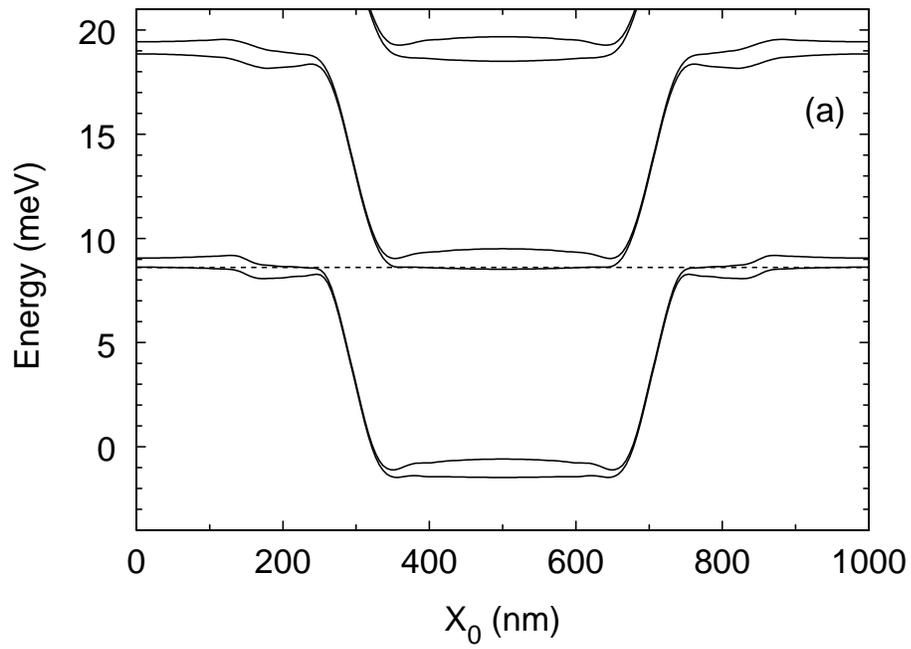
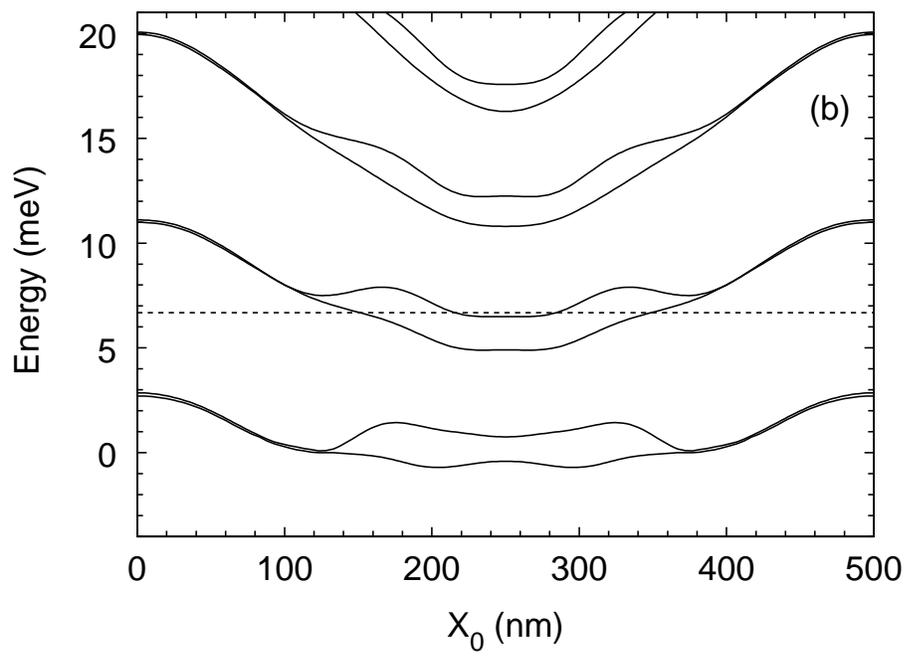
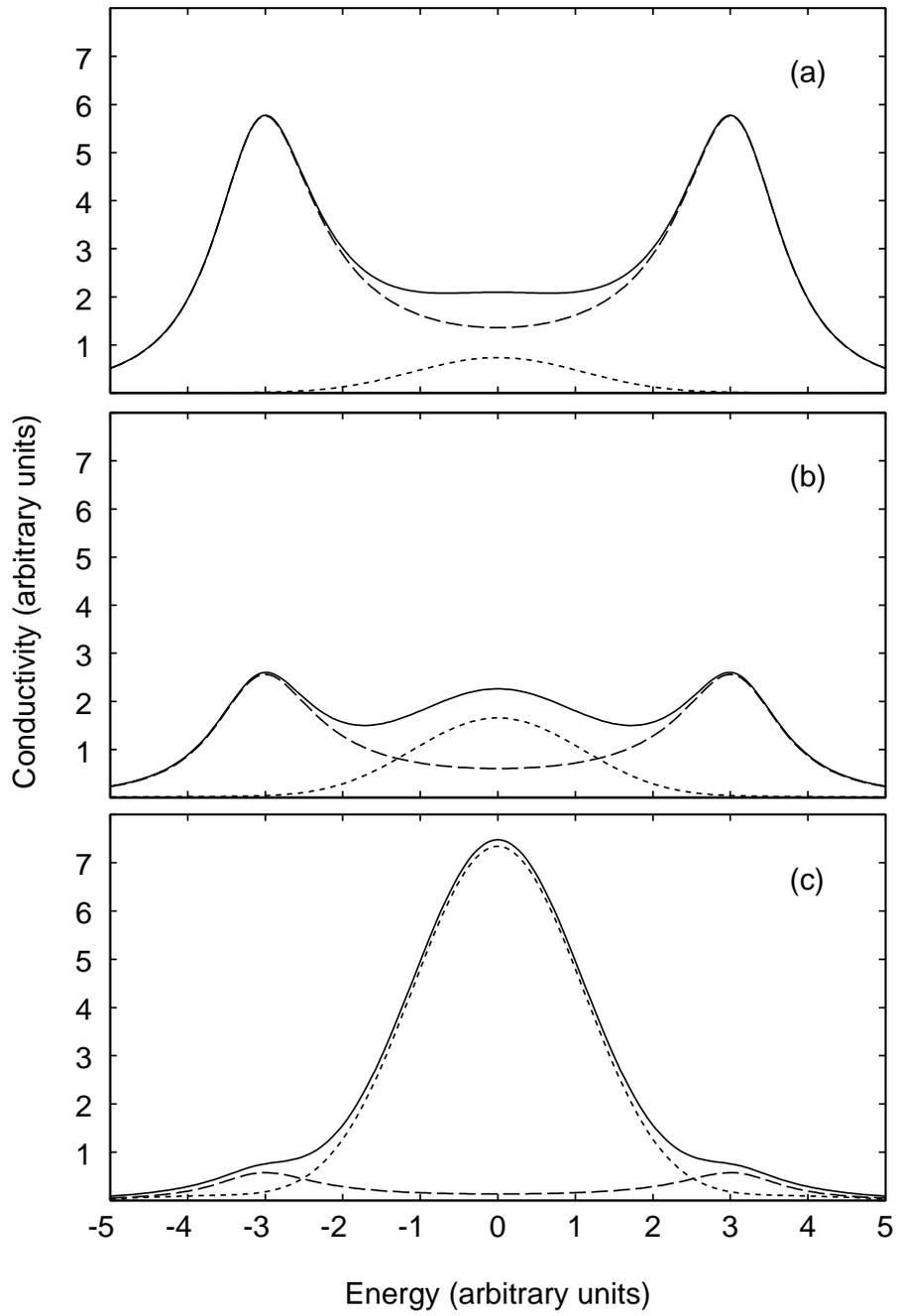
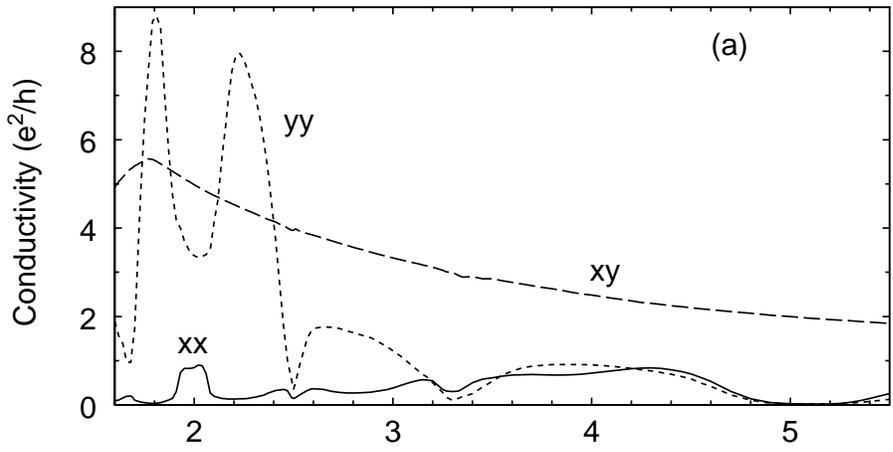
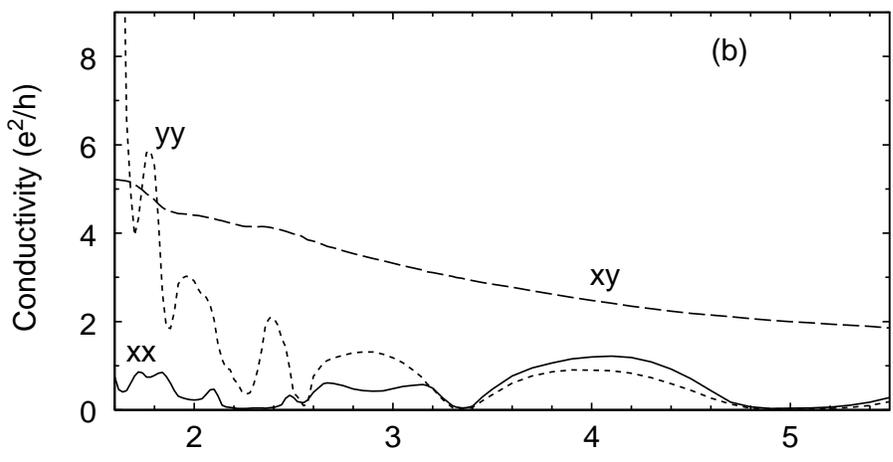
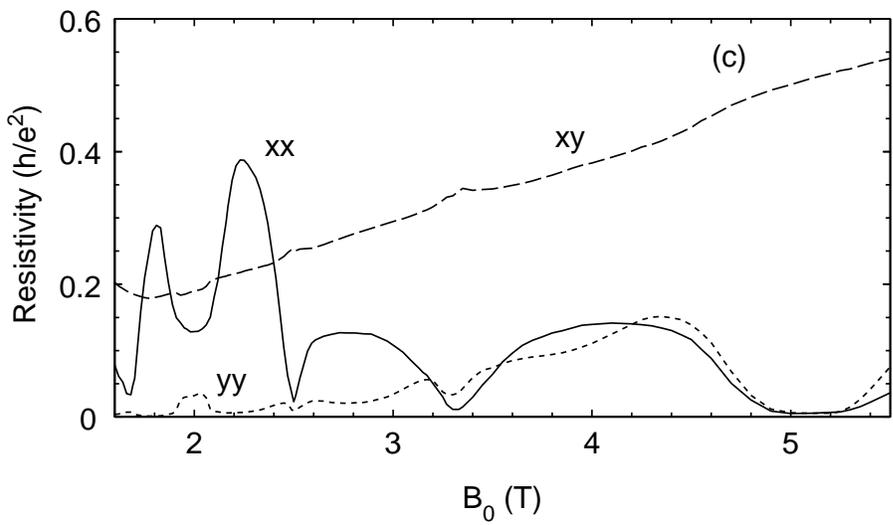
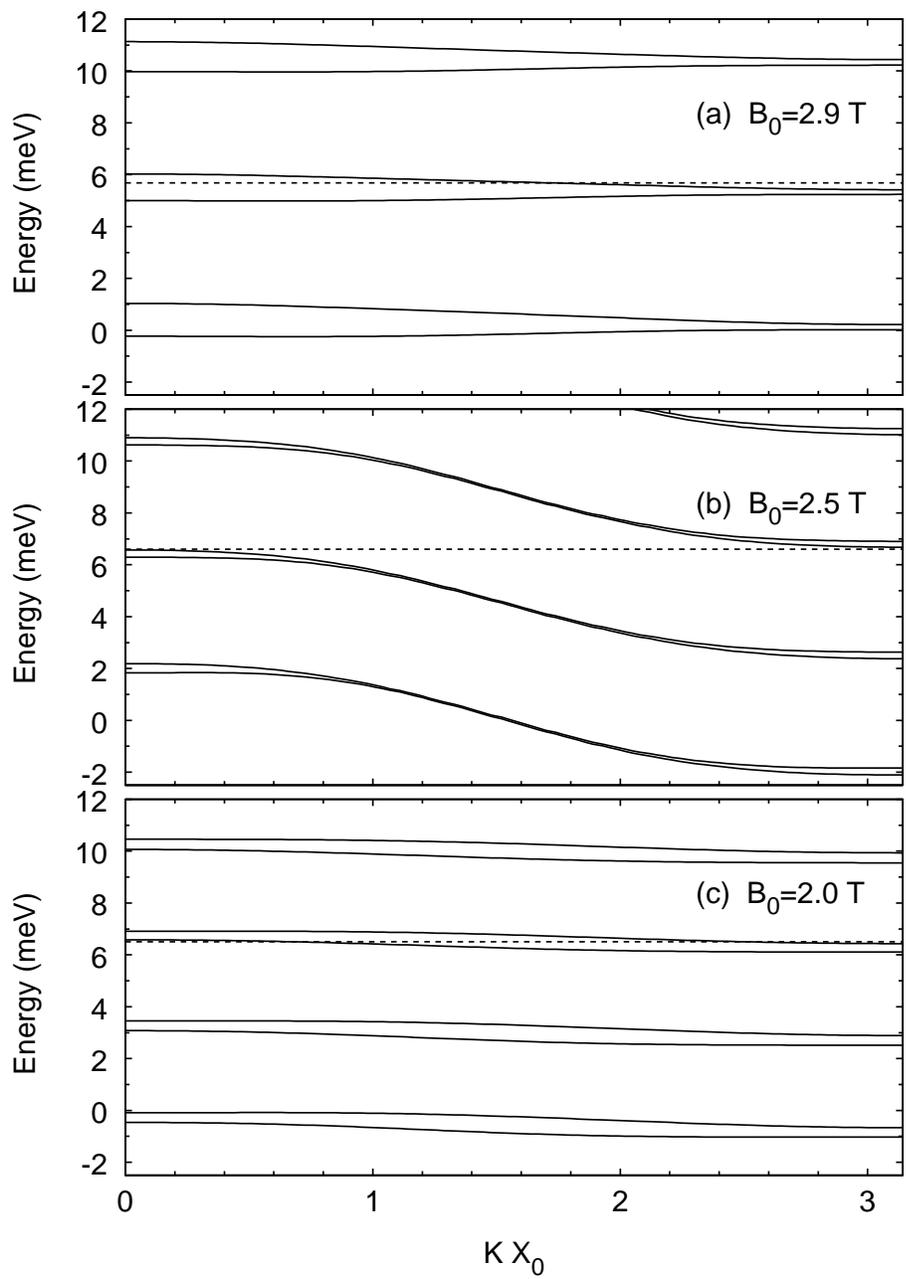
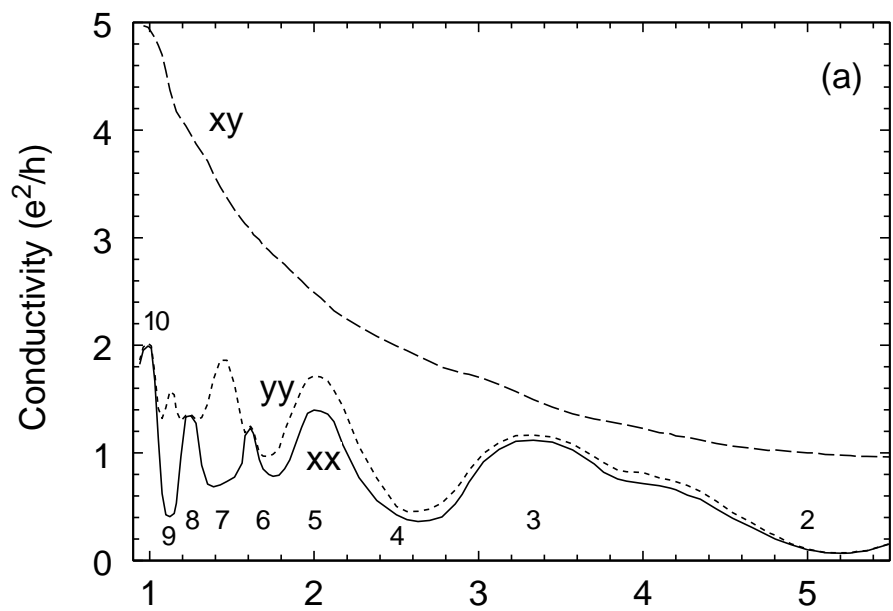
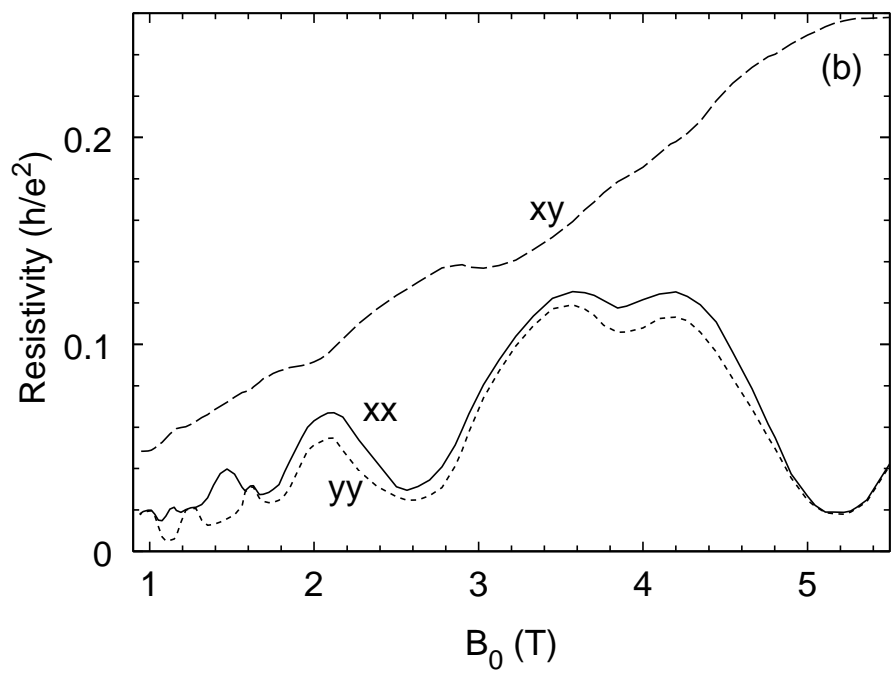
\end{document}